\documentstyle[12pt,aasms4,flushrt]{article}
\tighten
\slugcomment{To appear in {\it Astrophysical Journal Letters}}
\lefthead{DESCH \& ROBERGE}
\righthead{FAR-INFRARED POLARIZATION FROM THE GALACTIC CND}
\newcommand{\sameauthor}{\rule[+.5ex]{.7cm}{1pt}. \ }

\newcommand{\Konigl}{\mbox{K\"{o}nigl}}      
\newcommand{\mic}{\mbox{\,$\mu$m}}      
\newcommand{\vecJ}{\mbox{\boldmath$J$}}      
\newcommand{\vecB}{\mbox{\boldmath$B$}}      
\newcommand{\vecBav}{\mbox{\vecB$_{\rm avg}$}}      
\newcommand{\deltaB}{\mbox{\boldmath$\delta B$}}

\newcommand{\Bnu}{\mbox{$B_{\nu}(T_d)$}}      
\newcommand{\Cxp}{\mbox{$C_{x^{\prime}}$}}      
\newcommand{\Cyp}{\mbox{$C_{y^{\prime}}$}}      
\newcommand{\vecx}{\mbox{\boldmath$x$}}      
\newcommand{\vecz}{\mbox{\boldmath$z$}}      
\newcommand{\vecy}{\mbox{\boldmath$y$}}      
      
\newcommand{\pfxh}{\mbox{$\hat{\vecx}^{\prime}$}}
\newcommand{\pfyh}{\mbox{$\hat{\vecy}^{\prime}$}}
\newcommand{\pfzh}{\mbox{$\hat{\vecz}^{\prime}$}}

\newcommand{\cosbav}{\mbox{$\left<\cos^2\beta\right>$}}

\newcommand{\Cper}{\mbox{$C_{\perp}$}}      
\newcommand{\Cpar}{\mbox{$C_{\parallel}$}}      
\newcommand{\Cavg}{\mbox{$C_{\rm avg}$}}      
\newcommand{\vth}{\mbox{$v_{\rm th}$}}

\newcommand{\cmMMM}{\mbox{cm${}^{-3}$}}
\newcommand{\kms}{\mbox{km\ s${}^{-1}$}}

\newcommand{\nH}{\mbox{$n_{\rm H}$}}
\newcommand{\Pthr}{\mbox{$P_{\rm thr}$}}

\newcommand{\Pobs}{\mbox{$P_{\rm obs}$}}

\newcommand{\sigtrb}{\mbox{$\sigma_{{\rm trb}}$}}

\newcommand{\Td}{\mbox{$T_{\rm d}$}}
\newcommand{\nd}{\mbox{$n_{\rm d}$}}
\newcommand{\Tg}{\mbox{$T_{\rm g}$}}
\newcommand{\TdTg}{\mbox{$\Td/\Tg$}}

\newcommand{\cross}{\mbox{\boldmath$\times$}}      % cross product
\newcommand{\vd}{\mbox{$v_{\rm d}$}}
\begin{document}
%
%%%%%%%%%%%%%%%%%%%%%%%%%%%%%%%%%%%%%%%%%%%%%%%%%%%%%%%%%%%%%%%%%%%%%%%%%%%%
% Section 0: Title and abstract
%%%%%%%%%%%%%%%%%%%%%%%%%%%%%%%%%%%%%%%%%%%%%%%%%%%%%%%%%%%%%%%%%%%%%%%%%%%%
%
\title{Ambipolar Diffusion and Far-Infrared Polarization from the Galactic
Circumnuclear Disk}

\author{S. J. Desch\altaffilmark{1} and W. G. Roberge\altaffilmark{2}}
\altaffiltext{1}{Department of Physics, University of Illinois at Urbana-Champaign,
     1002 West Green St., Urbana, IL 61801, desch@astro.uiuc.edu}

\altaffiltext{2}{Dept.\ of Physics, Applied Physics \& Astronomy,
     Rensselaer Polytechnic Institute, Troy, NY 12180, roberw@rpi.edu}

\begin{abstract}
We describe an implicit prediction of the accretion disk
models constructed by Wardle and \Konigl\ (1990) 
for the circumnuclear disk (CND) of gas and
dust near the Galactic center:
supersonic ambipolar diffusion, an essential
dynamical ingredient of the Wardle-\Konigl\ disks,
will cause the alignment of dust grains due to a
process described by Roberge, Hanany, \& Messinger (1995).
We calculate synthetic maps of the polarized thermal emission which
would be caused by ambipolar alignment in the preferred
Wardle-\Konigl\ model.
Our maps are in reasonable agreement with
100\mic\ polarimetry of the CND if we assume that the grains have shapes
similar to those of grains in nearby molecular clouds and
that the CND contains a disordered magnetic field
in energy equipartition with its ordered field.
\end{abstract}

%
%%%%%%%%%%%%%%%%%%%%%%%%%%%%%%%%%%%%%%%%%%%%%%%%%%%%%%%%%%%%%%%%%%%%%%%%%%%%
% 1. Introduction
%%%%%%%%%%%%%%%%%%%%%%%%%%%%%%%%%%%%%%%%%%%%%%%%%%%%%%%%%%%%%%%%%%%%%%%%%%%%
%
\setcounter{footnote}{0}
\section{INTRODUCTION}

%
% Paragraph #I-1:
%  What is the CND; why it is interesting; what has been done already.
%

The Galactic circumnuclear disk (CND) is a weakly-ionized ring
of gas and dust which appears in various types of emission
at galactocentric radii $R=1$--$10$\,pc
(see Morris \& Serabyn 1996 and Davidson 1996 for recent reviews).
The origin and structure of the CND magnetic field, \vecB, are particularly
intriguing: mid- and far-infrared polarimetry
(\cite{AitkU86}; \cite{WernU88}; Hildebrand et al.\ 1990, 1993 [H90, H93])
show that \vecB\ lies primarily in the plane of the disk and is 
roughly orthogonal to the poloidal field observed on larger scales.
Werner et al. (1988) pointed out that a differentially-rotating disk
would deform an initially poloidal field into a predominantly toroidal
configuration.
Subsequently, Wardle and \Konigl\ (1990, [WK]) described self-similar
hydromagnetic accretion disk models of the CND; with the addition of a small
nonaxisymmetric distortion, the magnetic field predicted by
their preferred model, gc2, is in reasonable agreement with the field
inferred from far-infrared polarimetry (H93).

%
% Paragraph #I-2:
%  CND has big drifts; this leads to grain alignment; we will predict
%  FIR polarizations based on this; we will compare them with observations.
%

In the WK models, the magnetic torques which remove angular momentum
from the gas are mediated by supersonic ambipolar diffusion.
Coincidentally, Roberge, Hanany, \& Messinger (1995, [RHM]) showed
that ambipolar diffusion causes grain alignment:
the partial coupling of charged dust grains to the drifting
field lines produces gas-grain streaming, which causes
alignment of the grains via Gold's mechanism
(\cite{Gold52}; \cite{Purc69}; \cite{PS71}; \cite{Laza94}; RHM).
The large magnitude of the ``intrinsic'' 100\mic\ polarization
from grains aligned by supersonic ambipolar diffusion 
(e.g., $20$\% for oblate silicate grains with $2$:$1$ axis ratios viewed
with the most favorable magnetic field geometry; see RHM)
leads us to conjecture that the ambipolar mechanism is the
predominant cause of grain alignment in the CND.
In \S\S2--3, we test this conjecture by calculating the 
polarized thermal emission that would be caused by ambipolar alignment 
in model gc2 and comparing our results to the 100\mic\ polarimetry
of H90 and H93.

%
%%%%%%%%%%%%%%%%%%%%%%%%%%%%%%%%%%%%%%%%%%%%%%%%%%%%%%%%%%%%%%%%%%%%%%%%%%%%
% 2. Calculations
%%%%%%%%%%%%%%%%%%%%%%%%%%%%%%%%%%%%%%%%%%%%%%%%%%%%%%%%%%%%%%%%%%%%%%%%%%%%
%
\section{CALCULATIONS}

%
% Paragraph #II-1
%  Here are the equations used to predict polarizations
%
The largest optical depth through the CND is $< 0.1$ at 100\mic;
consequently the Stokes parameters, $I$, $Q$, and $U$, are well
described by the
transfer equations for optically thin emission,
\begin{equation}
{dI \over ds} = n_d\,\Bnu\,\left(\Cxp+\Cyp\right)
\label{eq-2_1}
\end{equation}
\begin{equation}
{dQ \over ds} = n_d\,\Bnu\,\left(\Cxp-\Cyp\right)\,F\,\cos 2\psi
\label{eq-2_2}
\end{equation}
\begin{equation}
{dU \over ds} = n_d\,\Bnu\,\left(\Cxp-\Cyp\right)\,F\,\sin 2\psi
\label{eq-2_3}
\end{equation}
(WK90), where $s$ is distance along the line of sight.
We integrated equations (\ref{eq-2_1})--(\ref{eq-2_3})
to obtain the polarization magnitude, $P=\sqrt{Q^2+U^2}/I$, and
position angle\footnote{With $\theta=0$ if the electric
vector is polarized along the east-west direction.},
$\theta=\frac{1}{2}\arctan(U/Q)$,
using a model of the position-dependent dust density, \nd,
and temperature, \Td, derived by Davidson et al.\ (1992)
from 50 and 90\mic\ photometry of the Galactic center.
Our results are only weakly sensitive to uncertainties in \nd\
and \Td\ because the CND is optically and geometrically thin
and the FIR emission is in the Rayleigh-Jeans limit.
The other variables in equations (\ref{eq-2_1})--(\ref{eq-2_3})
are determined by the alignment properties of the grains as follows.

% 
% Paragraph #II-2
%  Here are the parameters entering into the equations.
%
%To model the dust density, $n_d$, and temperature, $T_d$, we
%adopted the distributions derived by Davidson et al.\ (1992) from 50 
%and 90\mic\ photometry of the Galactic center.
%Thus, we represented the CND as a thin, flared disk with its
%rotation axis inclined by
%$i=60\arcdeg$ to the line of sight and its
%major axis, projected onto the sky, oriented $10\arcdeg$ east of north.
%We set $n_d=n_{{\rm d}0}\left(r/\rin\right)^{-1}$
%and $T_d=T_{{\rm d}0}\left(r/\rin\right)^{-1/3}$
%for $1$\,pc\,$<r<$\,6\,pc, where
%$\rin=1$\,pc and $T_{{\rm d}0}=75$\,K.\footnote{We have adjusted the
%linear dimensions given by Davidson et al.\ (1992)
%slightly to be consistent with
%the Sun-Galactic center distance adopted here.}
%For optically thin emission, the magnitude and position
%angle of the linear polarization are independent
%of $n_{{\rm d}0}$, whose value is therefore irrelevant.
%Because the disk is both optically and geometrically thin,
%our results are not very sensitive to uncertainties in $n_d$.
%Similarly, our results depend only weakly on $T_d$ because the
%latter varies only from $40$ to $75$\,K
%and the far-infrared emission is in the Rayleigh-Jeans limit.
%
%
% Paragraph #II-3:
%  Here is where the connection between polarization and grain orientation
%  is made.
%
The quantities \Cxp\ and \Cyp\ are respectively the
dust absorption cross sections for light polarized along
the \pfxh\ and \pfyh\ directions, where \pfyh\ is parallel
to the projection of the local magnetic field onto the plane of the sky,
\pfzh\ points toward the observer, $\pfxh=\pfyh\cross\pfzh$,
and $\psi$ is the angle between \pfyh\ and north,
measured counterclockwise from north.
To compute \Cxp\ and \Cyp, we modeled the grains as oblate
spheroids with semiaxes $a$ parallel to the symmetry axis 
and $b$ ($>a$) perpendicular to the symmetry axis.
We assumed that the angular momentum, \vecJ, of a rotating spheroid is
aligned perfectly with its symmetry axis by the Barnett
effect.\footnote{That is,
we neglected the disalignment of \vecJ\ by
thermal fluctuations (\cite{Laza94}).
This is a good approximation for the conditions
of interest here (Lazarian \& Roberge 1996).}
In this approximation,
\begin{equation}
C_{x^{\prime}} =
\Cavg + \frac{1}{3}Q_J\left(C_{\perp}-C_{\parallel}\right)
\label{eq-2_4}
\end{equation}
and
\begin{equation}
C_{y^{\prime}} =
\Cavg + \frac{1}{3}Q_J\left(1-3\cos^2\zeta\right)
\left(C_{\perp}-C_{\parallel}\right)
\label{eq-2_5}
\end{equation}
(\cite{LD85}), where
\Cpar\ and \Cper\ are respectively the cross sections for light
polarized parallel and perpendicular to the symmetry axis,
$\Cavg \equiv \left(2\Cper+\Cpar\right)/3$, and
$\zeta$ is the angle between \vecB\ and the plane of the sky.
For a given $a/b$ ratio,
we computed \Cpar\ and \Cper\ in the electric dipole approximation
using Draine's (1987) dielectric function for astronomical silicates.
Notice that $P$ and $\theta$ depend only on {\it ratios}\/ of the grain
cross sections, which are independent of grain size in
the electric dipole limit.
Thus, apart from its weak influence on the grain dynamics (discussed
below), the value of $b$ does not affect our calculations.
We assumed for simplicity that the grain shape is independent of
location and let $a/b$ be a free parameter.

%
% Paragraph #II-4
%  Here is where the connection between the polarizations and the 
%  magnetic field geometry is made.
%
The angles $\zeta$ and $\psi$ are functions of the position-dependent
disk magnetic field.
We computed $\zeta$ and $\psi$ from a numerical tabulation of model
gc2 kindly supplied by Mark Wardle.
The model is self-similar and to determine the hydro
variables uniquely it is also necessary to
specify the magnetic field, $B_0$, the density, $n_{{\rm H}0}$, and
the radial ($v_{r0}$) and azimuthal ($v_{\phi 0}$) velocity components
of the neutral gas at some fiducial radius, $r_0$.
We set $r_0=1$\,pc and chose $B_0=2$\,mG, consistent
with H\ I and OH Zeeman splitting measurements of the CND
(reviewed by Morris \& Serabyn 1996).
We set $n_{{\rm H}0}=10^5$\,\cmMMM, near the lower end of
the range of estimates for the density (\cite{JackU93}),
and assumed that $v_{\phi 0}=110$\,\kms\ and $v_r=-19$\,\kms,
consistent with kinematic models of the CND 
(\cite{GustU87}; \cite{JackU93}).

%
% Paragraph #II-5:
%  Here is where the connection to grain orientation (alignment) is made.
% 
The alignment of \vecJ\ with respect to \vecB\ is characterized by
the quantity
\begin{equation}
Q_J \equiv \frac{3}{2}
\left[\,\left<\cos^2\beta\right>-\frac{1}{3}\,\right],
\label{eq-2_6}
\end{equation}
where $\beta$ is the angle between \vecJ\ and \vecB\ and angle
brackets denote the average for all grains.
We calculated $Q_J$ from the theory of ambipolar
alignment (RHM).\footnote{Including
Davis-Greenstein alignment would increase $Q_J$,
and hence the polarizations, by an uncertain factor that depends
on poorly-known magnetic properties of the grains.
The increase would be $\lesssim 40$\% for grains composed
of ordinary paramagnetic substances but could be larger for
superparamagnetic grains.}
For supersonic drift speeds,
$Q_J$ depends only on $a/b$ and $\vd/\vth$, where \vd\ 
is the gas-grain drift speed and $\vth = \surd{2kT_{\rm g}/m}$ is the
gas thermal speed (RHM).
We computed \vd\ at each point using model gc2 to
predict the local hydro variables and
a prescription for the grain dynamics
(Draine 1980) which includes Lorentz and gas drag forces on the
grains.
For the purposes of calculating the grain dynamics only, we
represented the grains as spheres with equivalent radii
$r_{\rm eq} = \left(ab^2\right)^{1/3}$ and arbitrarily set $b=0.1$\mic.
Changing $b$ by a factor of 2 would have virtually no effect
on the calculated polarizations.
The WK models do not predict the gas temperature;
we set $\Tg=300$\,K,
consistent with observations of various molecular lines (\cite{HarrU85}).

%
% Paragraph #II-6:
%  Here is where the disorienting effects of a random magnetic field
%  are described.
%
The factor $F$ in equations (\ref{eq-2_2})--(\ref{eq-2_3})
represents the possible effects of a random magnetic field component.
We set the total field to $\vecBav+\deltaB$, where \vecBav\ is the
ordered field predicted by model gc2
and \deltaB\ is a random component.
We assumed that \deltaB\ is perpendicular to \vecBav\
(as would be the case if \deltaB\ is due to 
Alfv\'{e}nic turbulence) and that 
\deltaB\ has an axisymmetric, Gaussian distribution
of amplitudes (see \cite{MG91}).
For this model, it is straightforward to show that
\begin{equation}
F(\xi) =  {3 \over \sqrt{2\pi\xi} } \, \int_0^{\infty}\ 
     { e^{-x^2/2\xi} \, dx \over 1+x^2 }
     - \frac{1}{2},
\label{eq-2_7}
\end{equation}
where the ``turbulence parameter,''
$\xi$, is the ratio of the energies in the ordered and disordered field
components.
We assumed for simplicity that this ratio is independent of position and let
$\xi$ be a free parameter.
Note that $F$ decreases monotonically from
$F(0)=1$ (no disordered field) to $F(1) \approx  0.5$ (disordered
field in equipartition with the ordered field).

%
%%%%%%%%%%%%%%%%%%%%%%%%%%%%%%%%%%%%%%%%%%%%%%%%%%%%%%%%%%%%%%%%%%%%%%%%%%%%
% 3. Results
%%%%%%%%%%%%%%%%%%%%%%%%%%%%%%%%%%%%%%%%%%%%%%%%%%%%%%%%%%%%%%%%%%%%%%%%%%%%
%
\section{Results}

%
% Paragraph #III-1:
%  This is how we performed the calculations; these are the parameters;
%   this is what we hope to solve for or test.
%
We constructed synthetic maps of $\Pthr$, the 100\mic\ polarization predicted
by ambipolar alignment, for 3,136 parameter combinations
sampled uniformly on the intervals $0 < a/b < 1$ and $0 < \xi < 2$.
For each $\left(a/b,\xi\right)$ combination,
we calculated the Stokes parameters on a grid of sightlines
uniformly spaced by $1\arcsec$ in right ascension and
declination.
Each map of $I$, $Q$, and $U$ was then convolved
with a $45\arcsec$ (FWHM) Gaussian beam 
to model the observed resolution (H93)
and maps of \Pthr\ were computed from the convolved
Stokes parameters.
Optimal values of $a/b$ and $\xi$ were obtained
by performing a $\chi^2$ fit to \Pobs, the 100\mic\
polarization observed along $N=23$ lines of sight
(H90, H93).
% (Hildebrand et al.\ 1990, 1993).
Our fits omitted 7 observed
sightlines for which the FIR emission is not
associated with the CND.
We did not attempt to fit $\theta$, which is a good diagnostic
of the magnetic field geometry but not of the alignment mechanism.

%
% Paragraph #III-2:
%  Here we describe some statistical background explaining the reasons
%  why there will be a distribution of P around Pmean, some natural and
%  some due to inadequacies in the theory.
%
Strictly speaking, the value of \Pobs\ we calculate for each sightline
is the {\it mean}\/ polarization for a hypothetical ensemble of disks
with identical distribution functions for \deltaB.
That is, discrepancies between \Pthr\ and \Pobs\ are 
due not only to uncertainties in our theoretical model but also to
random fluctuations in \vecB.
To include the effects of random fluctuations on
$\chi^2$, we assumed that the polarization observed along the
$i$th sightline has dispersion
$\sigma_i^2 = \sigma_{{\rm obs},i}^2+\sigma_{{\rm trb}}^2$,
where $\sigma_{{\rm obs},i}$ is the observational uncertainty quoted
by H93 and \sigtrb\ represents the uncertainty due to fluctuations
in \vecB.
We assumed for simplicity that \sigtrb\ is the same for every
sightline and performed the $\chi^2$ fits independently for
different choices of \sigtrb.
We found that $\chi^2$ per degree of freedom is $\le 1$ in
some subset of parameter space if $\sigtrb \ge 0.6$\%.
That is, we are able to obtain a reasonable fit to the data if
we assume that the random fluctuations in \Pobs\ are about
half a percent.\footnote{This
logic probably overestimates $\sigtrb$, since
some of the errors that are ascribed to fluctuations in \Pobs\
are undoubtedly due to idealizations in our model.
For example, H93 pointed out that the WK models predict polarizations
that are symmetric in the northwest and southeast quadrants. However,
the observations clearly show departures from symmetry in the southeast
quadrant; not surprisingly, these points contribute about twice
as much on average to $\chi^2$ as the rest of the data.}

Figure~1 is a contour plot of $\chi^2$ per degree of freedom
($\chi^2/\nu$) for the case $\sigtrb=0.6$\%.
Evidently the observations are consistent with a continuum of
models with different combinations of $a/b$ and $\xi$; this
is due to the fact that a small increase in $a/b$
(which tends to reduce \Pthr) can always be compensated by
a decrease in $\xi$ (which tends to increase \Pthr).
Thus, the contours of constant $\chi^2/\nu$ slope downward with
increasing $a/b$.
It is interesting to note that the parameters allowed
by our fits are consistent with independent constraints:
%For example, Zeeman splitting observations of the 21\,cm HI (\cite{SL90})
%and 18\,cm OH (\cite{KLC92}) lines toward the Galactic center
%imply that the magnetic energy density of the CND
%is comparable to the turbulent energy density inferred from atomic and
%molecular linewidths (\cite{GustU87}; \cite{JackU93}).
%Thus, one expects $\xi\approx 1$ in the CND,
%consistent with observations of nearby molecular clouds
%(Jones 1989; Myers \& Goodman 1991) and the predictions of models for
%magnetic braking (Mouschovias \& Morton 1985).
For example, observations show nearby molecular clouds have
$\xi\approx 1$ (Jones 1989; Myers \& Goodman 1991).
Also, Zeeman splitting observations of the 21\,cm HI (\cite{SL90})
and 18\,cm OH (\cite{KLC92}) lines toward the Galactic center
imply that the magnetic energy density of the CND is comparable
to the turbulent energy density inferred from atomic and
molecular linewidths (\cite{GustU87}; \cite{JackU93}).
These observations are consistent with the presence of nonlinear
Alfv\'{e}n waves in the CND, for which $\xi\approx 1$.
Nonlinear Alfv\'{e}n waves are predicted by some models of magnetic braking
(Mouschovias \& Morton 1985).
Similarly, photopolarimetry of the 9.7\mic\ silicate
resonance implies that the aligned grains in nearby molecular
clouds are oblate with $a/b \approx 2/3$ (Hildebrand \& Dragovan 1995).
A model with $a/b=2/3$ and $\xi=1$ (indicated by the filled
circle in Fig.~1) is compared with the observations in Figure~2.
For this model, $\chi^2/\nu=0.97$ and the mean discrepancy
between \Pthr\ and \Pobs\ is $-0.2$\%.

%
% Paragraph #III-5:
%  Here we discuss the robustness of the results.
%
Our results are insensitive to the precise values of the
hydrodynamic variables; for example, the polarizations we
calculate would be virtually identical if we arbitrarily 
increased or decreased \nH\ or $B$ by factors of $\sim 3$.
This robustness is due to the fact that, for the highly
supersonic drift speeds in the WK models ($\vd \gtrsim 10\,\vth$ for gc2),
the efficiency of ambipolar alignment saturates at 
a value which depends only on the grain shape (see RHM, Fig. 12).
However, our results {\it are}\/ sensitive to the
magnetic field geometry. For example, we attempted to fit
the observations by arbitrarily replacing the magnetic field
predicted by model gc2 with a purely toroidal field of the
same magnitude at each point; we were unable to find solutions
with $\chi^2/\nu < 2$ for these models.
It will be interesting to see whether one can model the CND
observations successfully with more detailed hydro models (e.g., 
Wardle \& K\"{o}nigl 1993) which calculate the vertical structure of
the disk.

%
% Paragraph #III-6: 
%  The robustness of the results is related to the assumption made
%  elsewhere about "saturation", an assumption that proved correct.
%  This robustness is contrasted with the lack of such inherent in
%  D-G alignment.
%
Previous models of the CND magnetic field geometry (WK, H90, H93) have
accounted successfully for the observed polarization position angles
by assuming that the
efficiency of grain alignment is independent of position.
The saturation of ambipolar alignment at large drift speeds
provides a natural justification for this assumption.
In contrast, it is difficult to reconcile the uniform 
efficiency required by the observations with other alignment
mechanisms.
If the grains are superparamagnetic, then
the efficiency of Davis-Greenstein alignment would
be saturated but the saturation efficiency
would be very sensitive to the local dust-to-gas
temperature ratio:
independent calculations on DG alignment
(Lazarian 1995) show that
factor of 2 variations in $T_d/T_g$ would cause factor
of $\approx 3$ variations in \cosbav.
The efficiency of Purcell's mechanism (Purcell 1979) 
is insensitive to \TdTg\ but sensitive to the timescale
for changes in the grain surface properties; it is
difficult to see why the latter should be uniform throughout
the CND and have just the right value to reproduce the observations.
We conclude that the degree and uniformity of the alignment
required by the observations are natural consequences of ambipolar
alignment but difficult to explain with other mechanisms.

%
%%%%%%%%%%%%%%%%%%%%%%%%%%%%%%%%%%%%%%%%%%%%%%%%%%%%%%%%%%%%%%%%%%%%%%%%%%%%
% 4. SUMMARY
%%%%%%%%%%%%%%%%%%%%%%%%%%%%%%%%%%%%%%%%%%%%%%%%%%%%%%%%%%%%%%%%%%%%%%%%%%%%
%
\newpage
\section{SUMMARY}

\begin{enumerate}

\item
We have calculated the 100\mic\ linear
polarization from the CND using a model that
attributes the grain alignment to ambipolar diffusion
in the Wardle-\Konigl\ model accretion disks.
Our polarization maps depend on just 2 adjustable
parameters, the grain axis ratio, $a/b$, and
the mean ratio, $\xi$, of the energies in the random
and ordered magnetic field components.

\item
We estimated the values of $a/b$ and $\xi$ by performing
a $\chi^2$ fit to the observed 100\mic\ polarizations (H90, H93).
Our models provide a reasonable fit to the observations if we
assume that the $1\sigma$ fluctuations in the observed polarizations
due to turblence in the CND are $\ge 0.6$\%.

\item
The observations are consistent with a continuum of
models with different $\left(a/b,\xi\right)$ values.
A model with $a/b=2/3$ and $\xi=1$, the parameter values
implied by independent constraints, falls close to
the minimum of $\chi^2$ per degree of freedom.
For this model, $\chi^2/\nu=0.97$ and the mean error
in the predicted polarizations is $-0.2$\%.

\item
Our calculations imply that grains in the
CND must be aligned with an efficiency such that $\cosbav \approx 0.5$,
nearly independent of position.
The degree and uniformity of alignment required
by the observations are natural predictions of the ambipolar
mechanism but require ``fine tuning'' of the parameters
in other alignment theories.
\end{enumerate}

\acknowledgments

We thank Roger Hildebrand, Arieh K\"{o}nigl, and the anonymous
referee for helpful comments on earlier versions of the paper.
This work was partially supported by NASA grant NAGW-3001.

\clearpage

%
%%%%%%%%%%%%%%%%%%%%%%%%%%%%%%%%%%%%%%%%%%%%%%%%%%%%%%%%%%%%%%%%%%%%%%%%
%  Figure 1
%%%%%%%%%%%%%%%%%%%%%%%%%%%%%%%%%%%%%%%%%%%%%%%%%%%%%%%%%%%%%%%%%%%%%%%%
%
\newpage
\epsscale{0.80}
\begin{figure}
%\plotfiddle{sdwr97fig1.eps}{800pt}{0}{100}{100}{0}{0}
\plotone{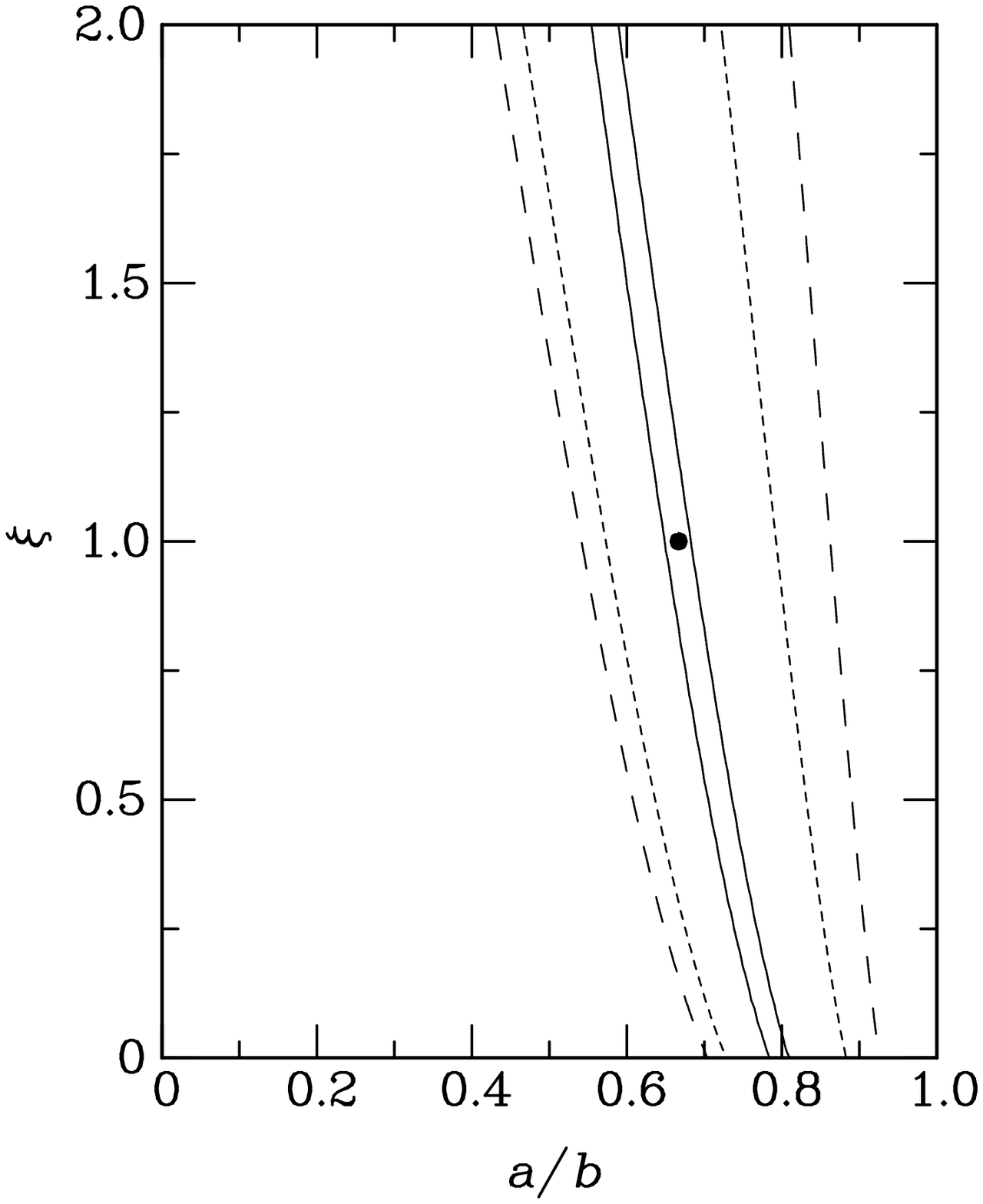}
\caption{
Contours of $\chi^2/\nu=1$ (solid lines), $2$ (short dash),
and $3$ (long dash) determined by fitting the grain axis
ratio, $a/b$, and turbulence parameter, $\xi$. The filled
circle marks the point
$\left(a/b,\xi\right)=\left(2/3,1\right)$
(see text).
}
\end{figure}

%
%%%%%%%%%%%%%%%%%%%%%%%%%%%%%%%%%%%%%%%%%%%%%%%%%%%%%%%%%%%%%%%%%%%%%%%%
%  Figure 2
%%%%%%%%%%%%%%%%%%%%%%%%%%%%%%%%%%%%%%%%%%%%%%%%%%%%%%%%%%%%%%%%%%%%%%%%
%
\newpage
\begin{figure}
\plotone{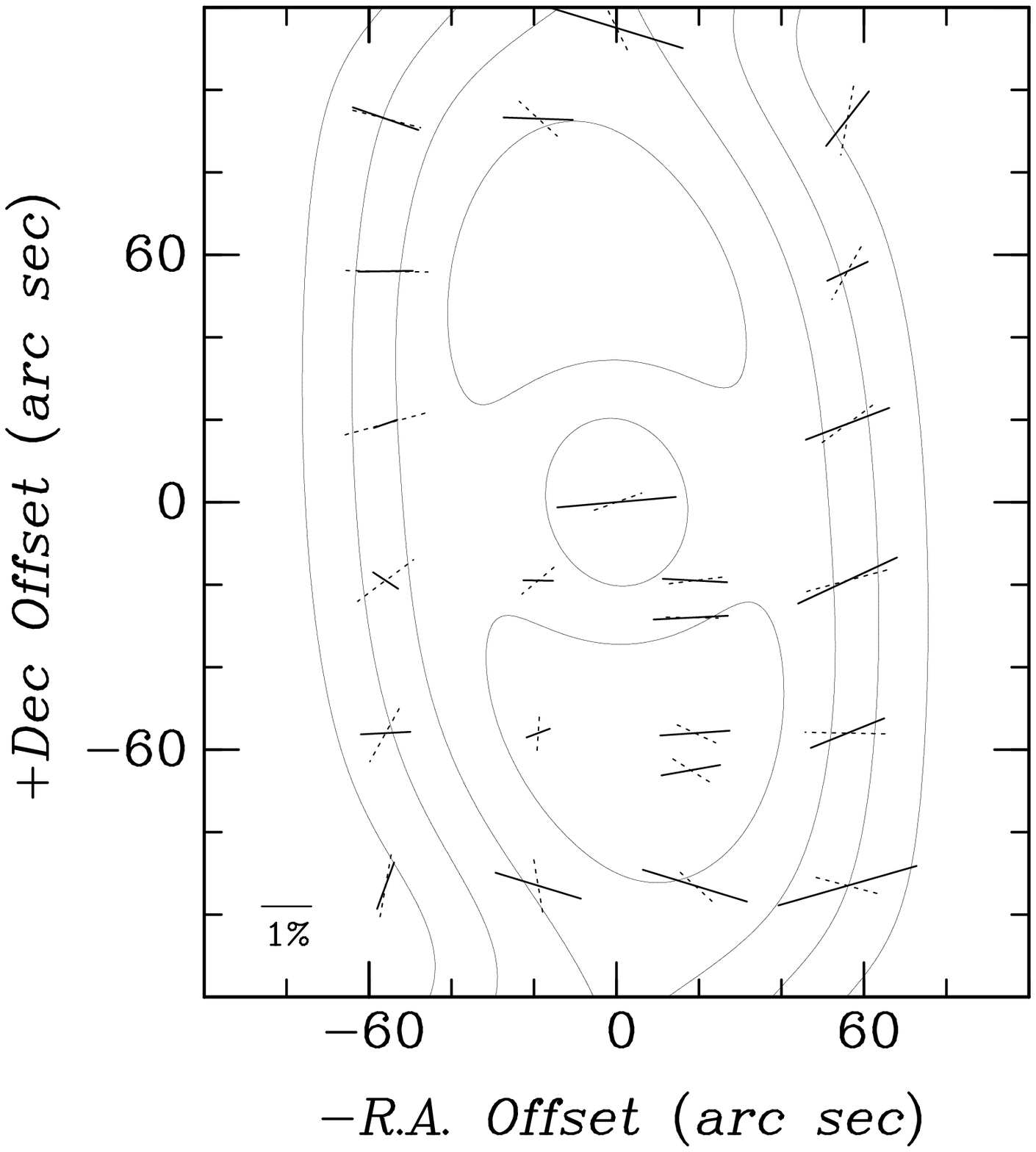}
\caption{
Comparison between the far-infrared polarizations observed
toward the CND (solid lines) and the predictions of a model
with $\left(a/b,\xi\right)=\left(2/3,1\right)$
(dashed lines).
Also shown are contours of the 100\mic\ intensity
predicted by our model, with contours at 20\%, 40\%, 60\%, and 80\%
of the peak intensity.
Offsets are measured from $\alpha=17^{\rm h}42^{\rm m}29\fs4$,
$\delta=-28\arcdeg 59\arcmin 19\arcsec$ (1950).
}
\end{figure}

%
%%%%%%%%%%%%%%%%%%%%%%%%%%%%%%%%%%%%%%%%%%%%%%%%%%%%%%%%%%%%%%%%%%%%%%%%%%%%
% End of manuscript
%%%%%%%%%%%%%%%%%%%%%%%%%%%%%%%%%%%%%%%%%%%%%%%%%%%%%%%%%%%%%%%%%%%%%%%%%%%%
%

\clearpage
\begin{thebibliography}{}
\bibitem[Aitken et al.\ 1986]{AitkU86}
     Aitken, D.K., Roche, P.F., Bailey, J.A., Briggs, G.P., Hough, J.H.,
     \& Thomas, J.A. 1986, \mnras, 218, 363
\bibitem[Davidson 1996]{Davi96}
     Davidson, J.A. 1996, in ASP Conf.\ Ser.\ Vol.\ 97, Polarimetry of
     the Interstellar Medium, ed. W. G. Roberge \& D. C. B. Whittet
     (San Francisco: ASP), 504
\bibitem[Davidson et al.\ 1992]{DaviU92}
     Davidson, J.A., et al.\ 1992, \apj, 387, 189
\bibitem[Draine 1980]{Drai80}
     Draine, B.T.\ 1980, \apj, 241, 1021
\bibitem[Draine 1987]{Drai87}
     \sameauthor 1987, Princeton Observatory Preprint 213
\bibitem[Gold 1952]{Gold52}
     Gold, T. 1952, \mnras, 112, 215
\bibitem[G\"{u}sten et al.\ 1987]{GustU87}
     G\"{u}sten, R., Genzel, R., Wright, M.C.H., Jaffe, D.T.,
     Stutzki, J., \& Harris, A.I. 1987, \apj, 318, 124
\bibitem[Harris et al.\ 1985]{HarrU85}
     Harris, A.I., Jaffe, D.T., Silber, M., \& Genzel, R. 1985,
     \apj, 294, L93
\bibitem[Hildebrand et al.\ 1990]{HildU90}
      Hildebrand, R.H., Gonatas, D.P., Platt, S.R., Wu, X.D.,
      Davidson, J.A., Werner, M.W., Novak, G., \& Morris, M.
      1990, \apj, 362, 114 (H90)
\bibitem[Hildebrand et al.\ 1993]{HildU93}
      Hildebrand, R.H., Davidson, J.A., Dotson, J., Figer, D.F.,
      Novak, G., Platt, S.R., \& Tao, L.
      1993, \apj, 417, 565 (H93)
\bibitem[Hildebrand \& Dragovan 1995]{HD95}
     Hildebrand, R.H., \& Dragovan, M. 1995, \apj, 450, 663
\bibitem[Jackson et al.\ 1993]{JackU93}
     Jackson, J.M., et al.\ 1993, \apj, 402, 173
\bibitem[Jones 1989]{J89}
    Jones, T.J. 1989, \apj, 346, 728
\bibitem[Killeen et al.\ 1993]{KLC92}
     Killeen, N.E.B, Lo, K.Y., \& Crutcher, R. 1992, \apj, 385, 585
\bibitem[Lazarian 1994]{Laza94}
     Lazarian, A. 1994, \mnras, 268, 713
\bibitem[Lazarian 1995]{Laza95}
     \sameauthor\ 1995, \apj, 453, 229
\bibitem[Lazarian \& Roberge 1996]{LR96}
     Lazarian, A., \& Roberge, W.G. 1996, \apj, submitted
\bibitem[Lee \& Draine 1985]{LD85}
     Lee, H.M., \& Draine, B.T.\ 1985, \apj, 290, 211
\bibitem[Morris \& Serabyn 1996]{MS96}
     Morris, M., \& Serabyn, E. 1996, \araa, 34, 645
\bibitem[Mouschovias \& Morton 1985]{MM85}
     Mouschovias, T. Ch., \& Morton, S.A. 1985, \apj, 298, 190
\bibitem[Myers \& Goodman 1991]{MG91}
     Myers, P.C., \& Goodman, A.A.\ 1991, \apj, 373, 509
\bibitem[Purcell 1969]{Purc69}
     Purcell, E.M.\ 1969, Physica, 41, 100
\bibitem[Purcell 1979]{Purc79}
     \sameauthor 1979, \apj, 231, 404
\bibitem[Purcell \& Spitzer 1971]{PS71}
     Purcell, E.M., \& Spitzer, L., Jr.\ 1971, \apj, 167, 31 
\bibitem[RHM]{RHM95}
     Roberge, W.G., Hanany, S., \& Messinger, D.W.\ 1995,
     \apj, 453, 238 (RHM)
\bibitem[Schwarz \& Lazenby 1990]{SL90}
     Schwarz, U.L., \& Lazenby, J. 1990, in
     Galactic and Extragalactic Magnetic Fields,
     eds.\ R. Beck, P. Kronberg, and R. Wielebinski, p. 383
\bibitem[WK90]{WK90}
     Wardle, M., \&  K\"{o}nigl, A. 1990, \apj, 362, 120 (WK)
\bibitem[Wardle \& K\"{o}nigl 1993]{WK93}
     \sameauthor 1993, \apj, 410, 218
\bibitem[Werner et al.\ 1988]{WernU88}
      Werner, M.W., Davidson, J.A., Morris, M., Novak, G.,
      Platt, S.R., \& Hildebrand, R.H.\ 1988, \apj, 333, 729
\end{thebibliography}
\end{document}